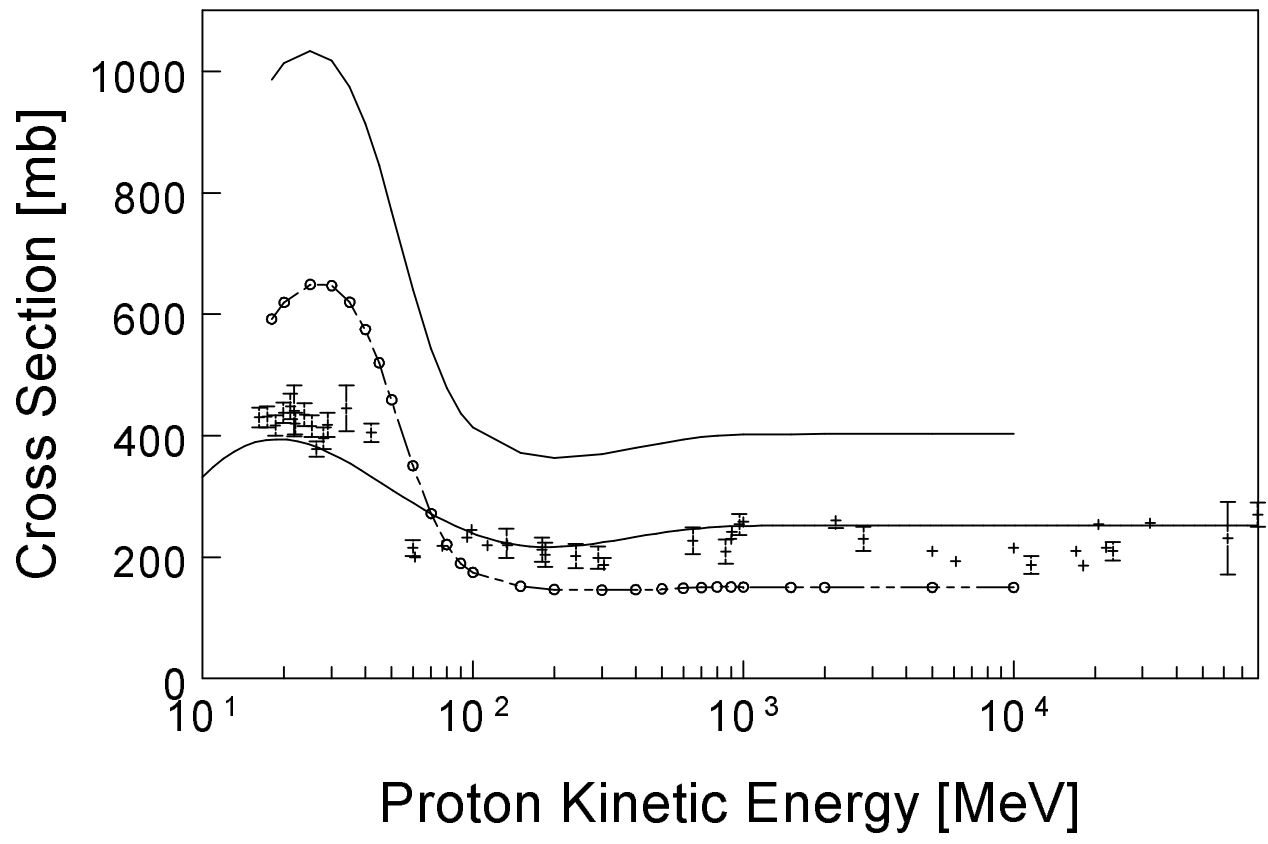

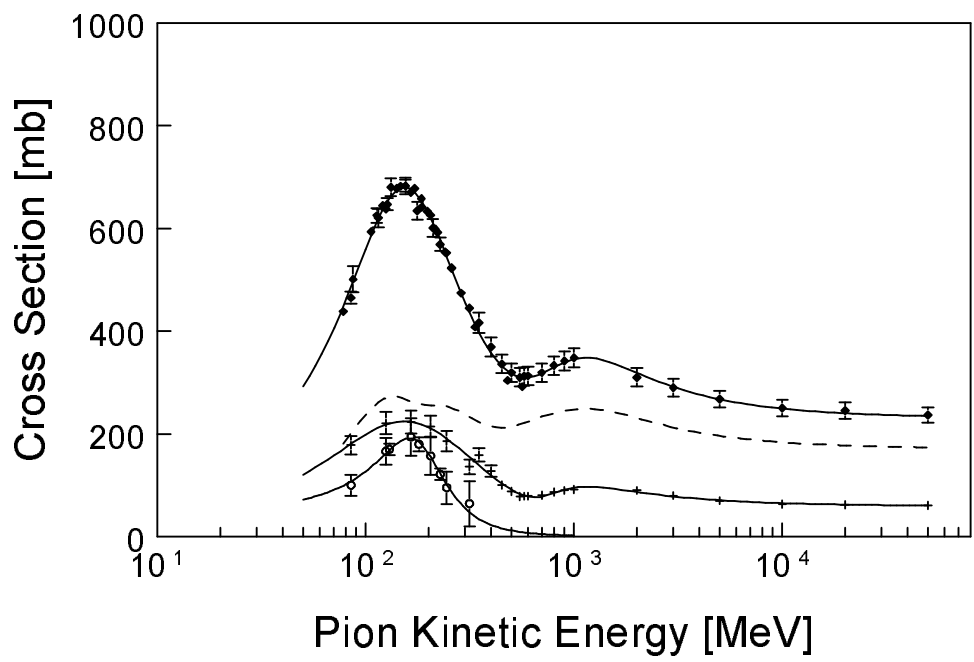

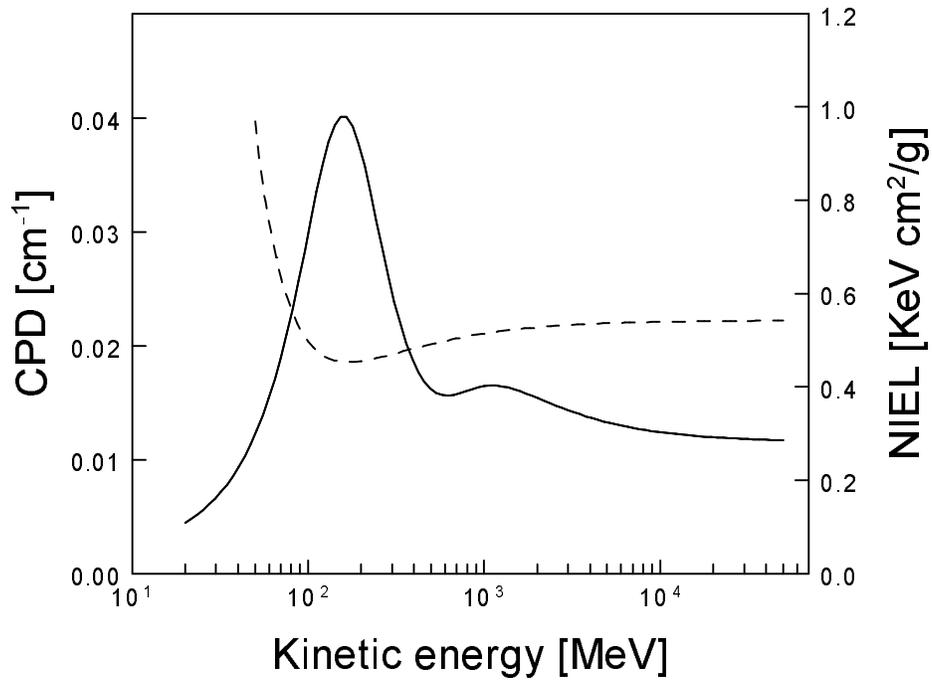

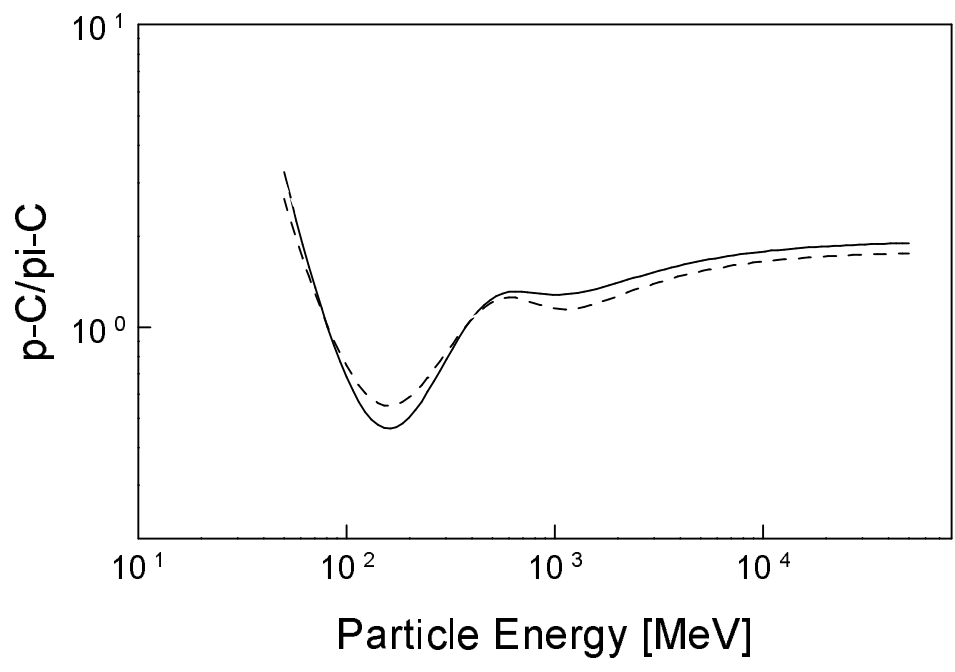

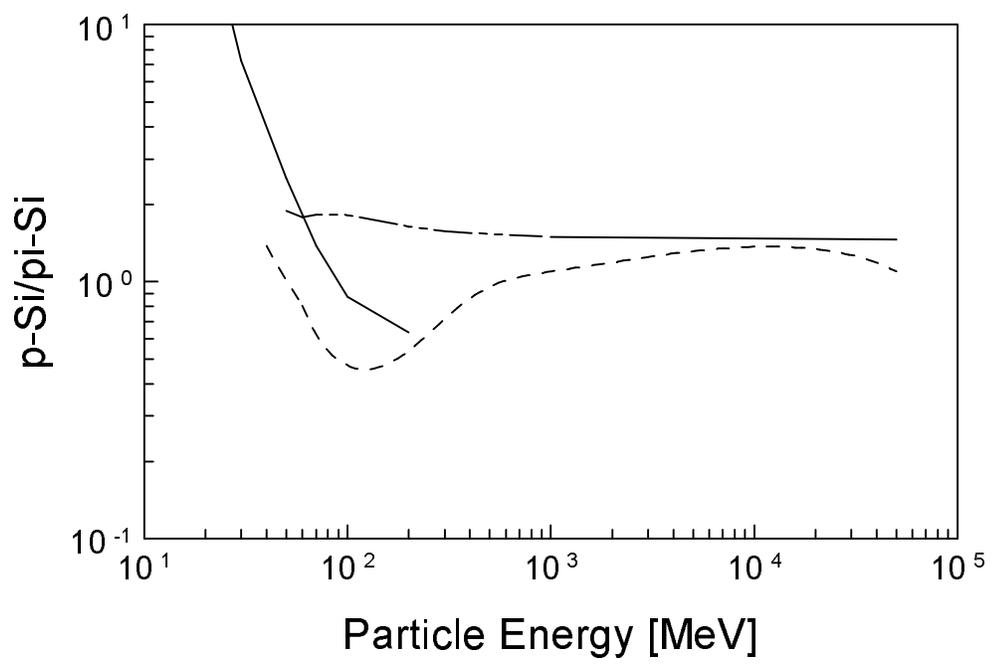

# Comparative Energy Dependence of Proton and Pion Degradation in Diamond[1]

*I.Lazanu and S.Lazanu[+]*

University of Bucharest, Faculty of Physics, POBox MG-11, Bucharest, Romania
[+]National Institute of Materials Physics, POBox MG-7, Bucharest-Magurele, Romania

**Abstract**

A comparative theoretical study of the damages produced by protons and pions, in the energy range 50 MeV ÷ 50 GeV, in diamond, is presented. The concentration of primary defects (CPD) induced by hadron irradiation is used to describe material degradation. The CPD has very different behaviours for protons and pions: the proton degradation is important at low energies and is higher than the pion one in the whole energy range investigated, with the exception of the $\Delta_{33}$ resonance region, where a large maximum of the degradation exists for pions.

In comparison with silicon, the most investigated and the most utilised semiconductor material for detectors, diamond theoretically proves to be one order of magnitude more resistant both to proton and to pion irradiation.

---

[1] accepted for publication in *Nuclear Instruments and Methods in Physics Research A* (1999)

# Comparative Energy Dependence of Proton and Pion Degradation in Diamond

*I.Lazanu and S.Lazanu\**

**1. Introduction**

Diamond shown promising properties [1] for its use as a very fast position sensitive detector for experiments in the highest radiation levels at the large hadron collider.

Up to now, the radiation resistance of diamond detectors has been demonstrated for photons and electrons, and experimental studies for pion, proton and neutron fields are in progress [2].

The theoretical calculations of diamond damage by $\pi^+$ and $\pi^-$ mesons in the $\Delta_{33}$ resonance energy range have been reported in reference [3], and the extension up to 50 GeV pion kinetic energy in reference [4], respectively.

For proton irradiation, some results exist for silicon [5,6,7], GaAs [8,9], InP [9], Ge [10], and also for diamond [11].

In this work, the theoretical estimation of the degradation of diamond in proton fields is presented comparatively with the similar calculations for pions, in the energy range 50 MeV - 50 GeV.

The physical quantity relevant to characterise the material degradation in radiation fields is the concentration of primary defects (CPD) produced by hadrons in the diamond lattice. For monoelement materials, the CPD is proportional to the non-ionising energy loss (NIEL), historically used to characterise the lattice degradation in particle fields. The CPD better correlates the damages produced in different materials at the same kinetic energy of the incident hadron.

**2. Model calculations for the degradation**

The CPD has been calculated as:

$$CPD = \frac{n(E)}{\Phi(E)}$$

with:

$$n(E) = \frac{N}{2E_d} \Phi(E) \sum_k \int d\Omega \sum_i \frac{d\sigma_i}{d\Omega} L(E_{Ri})$$

where: $E$ is the kinetic energy of the incident hadron; $N$ - atomic density of the target material; $E_d$ - threshold energy for displacements in the lattice; $\Phi(E)$ - the pion fluence in the primary beam; $E_{Ri}$ - recoil energy of the residual nucleus form the interaction mechanism $i$, from the interaction $k$ ($k=$ elastic, absorption and inelastic if the hadron is a pion, and $k =$ elastic and inelastic if it is a nucleon, respectively), having a $d\sigma_i / d\Omega$ - differential cross section; $L(E_{Ri})$ - Lindhard factor describing the partition between ionisation and displacements. The energy channelled into displacements, for each recoil (characterised by its charge and mass number), and for each energy, has been taken from reference [3].

The contribution of each channel to the total concentration of defects depends on the probability of interaction and on the kinematics of the process, reflected in the recoil energy of the residual nucleus.

In an elastic scattering process of interaction, symbolically represented by:

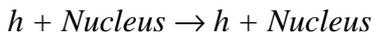

$h + Nucleus \rightarrow h + Nucleus$

the hadron does not excite the target nucleus.

The inelastic hadron - nucleus scattering includes all reactions of the type:

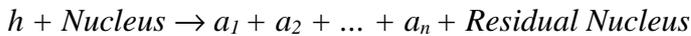

$h + Nucleus \rightarrow a_1 + a_2 + ... + a_n + Residual\ Nucleus$

where the reaction products $a_1, a_2, ..., a_n$ can be pion, proton, neutron, deuteron, other particles or light nuclei.

When the kinetic energy of the incident particle exceeds the threshold energy of 140 MeV, secondary pions could also be produced.

If the inelastic process is produced by nucleons, the identity of the incoming projectile is lost, and the creation of the secondary particles is associated with energy exchanges which are of the order of MeV or larger.

For pion - nucleus processes, a characteristic interaction is possible: the absorption, the process by which the pion disappear as a real particle within the nucleus. In these calculations, the absorption is considered separately. Absorption on a single nucleon is kinematically prohibited, and the simplest process is on two nucleons. Absorption on more nucleons is also possible.

The interaction of pions with nucleons and nuclei at kinetic energies comparable to the pion rest mass is dominated by the delta resonance production, with spin and isospin 3/2. At higher energies, other resonances could be produced, but with much less importance, and the pion behaviour does not differ significantly by that of other hadrons.

The energy dependence of cross sections, for proton and pion interactions with the carbon nucleus, present very different behaviours: the proton - nucleus cross section decreases with the increase of the projectile energy, then has a minimum at relatively low energies, followed by a smooth increase, while the pion - nucleus cross sections present for all processes a large maximum, at about 160 MeV, and which reflects the resonant structure of the interaction.

Since the inelastic and absorption collisions are considerably more complex than elastic scattering, special reaction models must be developed for their analysis.

For the concrete calculations, the available experimental data have been used, and also different phenomenological approximations.

In Figures 1 and 2, the energy dependencies of the proton - carbon and pion - carbon cross sections are shown, respectively.

For proton - carbon interactions, the elastic cross sections are from reference [12], the inelastic ones (data) from reference [13], and the continuous curve represent the parametrisation from reference [14].

For positive pions - carbon interactions, the data are from reference [15] for the elastic cross sections, from [16,17,18] for reaction cross sections, and from [17,19] for absorption, respectively. The continuous curves are best fits of the data, and have been used to extrapolate / interpolate the values of the cross sections at energies of interest. The inelastic cross sections are obtained as differences between reaction and absorption ones, and are represented as dashed lines in Fig. 2.

Elastic and Rutherford contributions to the CPD have been treated together, and the existent data for differential cross sections have been extrapolated at other energies in the frame a simple optical model.

The main difficulty is related to the inelastic interaction, due to the multitude of open channels, corresponding to possible final states. Some simplifying assumptions concerning this interactions have been made. Both for pions and protons, the knock-out interaction has been considered separately, using the data from [17,20,21] for pions, and from [12] for protons, respectively. The rest of the channels have been considered as equivalent to the interaction on an effective number of nucleons. Particle generation has been neglected.

In the case of pions, in the energy range of the delta resonance, another important contribution is given by absorption; above 1 GeV this process is negligible. Details about the contributions of different absorption mechanisms could be found in [2].

### 3. Results and interpretation

The results obtained for the energy dependence of the CPD (and NIEL) produced by protons and pions are represented on the same graph in Figure 3.

The CPD for protons present an abrupt decrease at low energies, followed by a minimum and, at higher energies, by a plateau. For pions, there exists a large maximum in the region of the $\Delta_{33}$ resonance. The minimum for proton degradation and respectively the maximum for pion one are in the same energetic range.

For both protons and pions, the energy dependence of the CPD follows mainly the energy dependence of the cross sections. The value of CPD at minimum for protons and the corresponding maximum for pions are in the ratio of approximately 0.47, while at the highest energies where calculations were performed, the ratio is around 1.95.

In Figure 4, the ratio of CPD produced by protons and pions in diamond (continuous line) is represented on the same graph with the corresponding ratio of the proton and pion total cross sections in carbon (dashed line). The two curves have a similar shape: a decreasing dependence from low energies up to the region corresponding to the delta resonance in the pion interaction, and a smooth energy dependence at higher energies with a minimum around 1 GeV; at high energies, the ratio of the CPD for protons and pions in diamond is proportional to the corresponding ratio of the total cross sections. This behaviour is in agreement with the estimations of Tang [22] on the distribution of secondary fragments resulting from the interaction.

For comparison, similar curves have been represented for silicon in Figure 5. There exist some NIEL calculations for protons, reported in references [5] and [6], and for pions: [4,5] and [23]. For the pion calculations, the results of Van Ginneken are accurate especially at high energies, because the resonant behaviour has been considered only by an increase of the cross sections and not as a specific mechanism of interaction. The Huhtinen and Aarnio calculations for pions are simple estimations, obtained using a scaling procedure of the proton NIEL results. At low energies, the ratio of the two degradations is estimated from references [6] and [4] (continuous line) and from reference [5] (continuous line for the energies where the results are accurate and dash - dotted line for the rest). The ratio of the total cross sections is represented with dashed line. The results suggest the similarity between diamond and silicon from the point of view of the ratio of degradations produced by protons and pions, and suggest also the possibility of applying a scaling procedure at energies higher then the delta resonance.

As absolute values, from the comparison of the present results for proton and pion degradation of diamond with the corresponding ones in silicon, the diamond proves to be one order of magnitude more resistant both to proton and to pion irradiation.

## 4. Summary

A comparative theoretical study of the damages produced by protons and pions in diamond has been done in the energy range between 50 MeV and 50 GeV. The concentration of primary defects in the diamond lattice was used to describe the radiation effects.

The CPD produced by protons and pions have very different energy dependencies. The CPD for protons presents an abrupt decrease at low energies, followed by a minimum and, at higher energies, by a plateau. For pions, there exists a large maximum in the region of the $\Delta_{33}$ resonance. The minimum for proton degradation and respectively the maximum for pion one are in the same energetic range. For both protons and pions, the energy dependence of the CPD follows mainly the energy dependence of the total cross sections.

In comparison with silicon, the most investigated and the most used semiconductor material for detectors, diamond theoretically proves to be one order of magnitude more resistant to proton and pion irradiation. Experimental studies are necessary to confirm these results.

# Figure captions

Figure 1. Energy dependence of proton - carbon cross sections: total, elastic and inelastic. The continuous curves represent parametrisations of the data.

Figure 2. Energy dependence of pion - carbon cross sections: up to down the curves represent total, inelastic, elastic and absorption cross sections. The curves are the best fits of the data.

Figure 3. Concentration of radiation induced defects (left scale) and non-ionising energy loss (right scale) produced by protons (dashed line) and pions (continuous line) in diamond.

Figure 4. Ratio of proton - carbon and pion - carbon total cross sections (dashed line) and ratio of CPD produced by protons and pions in diamond (continuous line), versus particle kinetic energy.

Figure 5. Energy dependence of the ratio of proton to pion silicon total cross sections (dashed line) and energy dependence of the ratio of the degradation produced by protons and pions in silicon as follows:
   - proton degradation from reference [6],and pion one from ref. [4] (continuous line);
   - proton and pion calculation after reference [5]: dashed - dotted between 50 MeV and 1 GeV, and continuous between 1 and 50 GeV.